\documentclass[conference, a4paper]{IEEEtran}
\IEEEoverridecommandlockouts
\usepackage{cite}
\usepackage{amsmath,amssymb,amsfonts}
\usepackage{graphicx}
\usepackage{textcomp}
\usepackage{xcolor}
\usepackage{algorithm}
\usepackage{algpseudocode}
\usepackage{amsmath}
\usepackage{graphics}
\usepackage{epsfig}
\usepackage{float}
\usepackage{bm}
\def\BibTeX{{\rm B\kern-.05em{\sc i\kern-.025em b}\kern-.08em
    T\kern-.1667em\lower.7ex\hbox{E}\kern-.125emX}}
\begin{document}

\title{Multiple Intelligent Reflecting Surface aided Multi-user Weighted Sum-Rate Maximization using Manifold Optimization\\
\author{\IEEEauthorblockN{Liyue Zhang\textsuperscript{1}, Qing~Wang\textsuperscript{1}, Haozhi Wang\textsuperscript{1}}
	\IEEEauthorblockA{\textsuperscript{1}Tianjin University, Tianjin, China \\
		Email: \{zhangliyue, wangq, wanghaozhi\}@tju.edu.cn}
}\thanks {This paper was published in the 10th IEEE/CIC International Conference on Communications in China (ICCC2021), held in Xiamen, China, 28-30 July 2021.}
}

\maketitle

\begin{abstract}
Intelligent reflecting surface (IRS) are able to amend radio propagation condition tasks on account of its functional properties in phase shift optimizing.
In fact, there exists geometry manifold in the base-station (BS) beamforming matrix and IRS reflection vector. Therefore, we propose a novel double manifold alternating optimization (DMAO) algorithm which makes use of differential geometry theory to improve optimization performance. First, we develop a multi-IRS multi-user system model to maximize the weighted sum-rate, which may lead to the non-convexity in our optimization procedure. Then in order to allocate an optimized coefficients to each BS antenna and IRS reflecting element, we present the beamforming matrix and reflection vector using complex sphere manifold and complex oblique manifold, respectively, which integrates the inner geometry structure and the constrains. 
By an innovative alternative iteration method, the system gradually converges an optimized stable state, which is associating with the maximized sum-rate. Furthermore, we quantize the IRS reflection coefficient considering the practical constrains. Experimental results demonstrated that our
approach significantly outperforms the conventional methods in terms of weighted sum-rate.
\end{abstract}


\section{Introduction}
Massive MIMO and millimeter wave will be applied to 5G networks \cite{5G}. However, the large-scale antenna arrays and mmWave frequency bands have high power consumption and high hardware cost \cite{myth}\cite{scene}. IRS is a 2-D planar surface composed of reconfigurable and near-passive reflecting elements with less energy drive, each of them being able to independently adjust the amplitude and phase of the incident signal. This advantage make IRS widely applied to reduce energy consumption and improve the propagation conditions \cite{towards}\cite{MIMO1}.  
By optimizing the phase-shift of each IRS reflecting element and the beamforming matrix of the BS according to the perfect CSI, we can reduce the transmit power \cite{li2019joint}\cite{wu2020joint}, extend the wireless coverage \cite{scene}, improve physical layer security \cite{cui2019secure}\cite{xu2019resource} and improve the channel capacity \cite{scene}\cite{yu2019miso}-\cite{chen2019sum}. 

Considering the channel capacity improving problem, the authors in \cite{yu2019miso} considered the scenario where there is only one user in wireless communication. In \cite{guo2019weighted}, the authors considered a multi-user scenario and maximized the weighted sum-rate of all users but there is only one IRS in the system. These method only treat one specific scenario. To this end, we aim to model the muti-user multi-IRS optimization problem, which can treat the single-user and single-IRS problem as the special cases.

Recently, iterative optimization methods have been successfully
applied in IRS capacity improving problem, such as local search and cross-entropy based algorithm \cite{chen2019sum}. These methods usually make use greedy search with high computational complexity.  
Author in \cite{scene} transform the problem into convexity case, which can not be extended to more general problem. Thus, we are motivated to explore non-convexity optimization methods to maximize the overall weighted sum-rate with low computational complexity.

\textbf{Why manifold optimization}:
Recent years we have witnessed the great development of Riemannian optimization algorithm in various types of matrix manifolds. In \cite{li2020manifold}, so as to joint design transmit waveform and receive filter for MIMO-STAP radars, the authors constructed the constraint into a smooth and compact manifold to reduce the computational cost. In the IRS-aided MIMO system, the authors in \cite{lin2020channel} proposed a channel estimation algorithm by leveraging fixed-rank manifold. The authors in \cite{hong2020semi} used sphere manifold for data detection. 
Therefore, 
taking the geometrical structures and the low-dimension feature of the manifold, we can solve the non-convexity problem with high efficiency when the feasible sets have manifold features, especially in high dimensions \cite{douik2019manifold}.

In this paper, we propose a double manifold alternating optimization (DMAO) algorithm to maximize the weighted sum-rate in IRS-aided multi-user MISO system with inter-user interference, by optimizing the beamforming matrix of the BS and the reflection coefficient of each IRS element. Specifically, we use the structural information of the constraint and replace the beamforming matrix and phase-shift vector using complex sphere and complex oblique manifolds, respectively. Via alternatively iteration using geometric conjugate gradient method, we improve the channel capacity significantly.

\textit{Notations}: Scalars are denoted by italic letters, vectors and matrices are denoted by bold-face
lower-case and upper-case letters, respectively. $ \mathbb{C}^{n\times p} $ denotes the space of $ n\times p $ complex-valued matrices. $ \mathrm{diag}\left(\cdot \right)  $ denotes the diagonal operation and $ \mathrm{vec}\left(\cdot \right)$ is the inverse operation of $ \mathrm{diag}\left(\cdot \right)  $. $ \mathrm{ddiag}\left(\cdot \right)  $ is a operation that sets all off-diagonal entries of a matrix to zero. $ \mathbb{E}\{\cdot\} $  is the expectation operator. $ \otimes $ is the Kronecker product. $ (\cdot) ^{*} $, $ (\cdot)^{T} $ and $(\cdot)^{H}$ denote conjugate, transpose, and conjugate transpose operations, respectively.
\section{System Model and Problem Formulation}\label{2}

\begin{figure}[ht]
	\centerline{\includegraphics[width=0.34\textwidth]{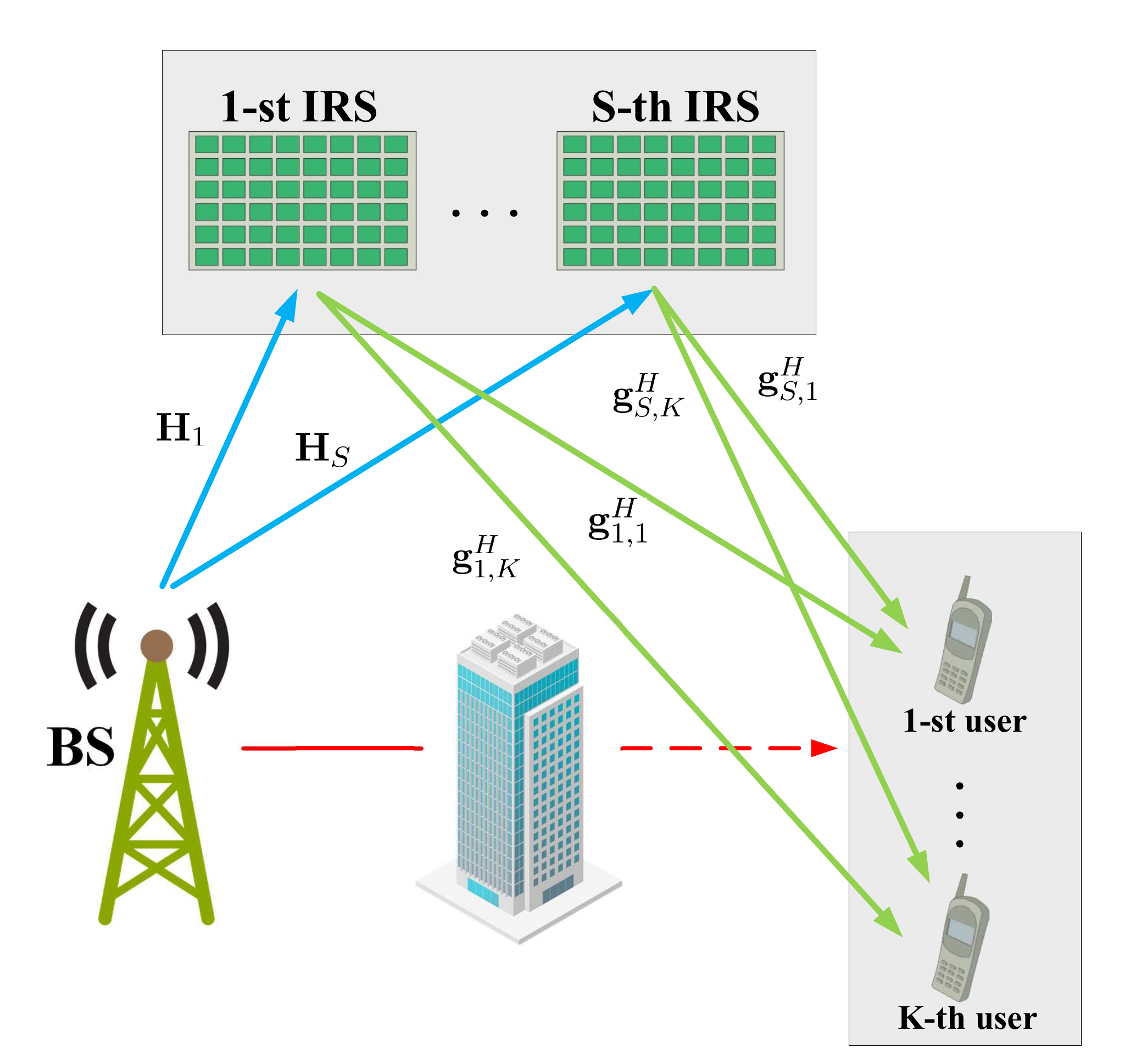}}
	\caption{IRS-aided multiuser MISO system.}
	\label{v}
\end{figure}

We consider an IRS-aided MISO system  as shown in Fig. \ref{v}, where the $ K $ users equipped with single receive antenna are served by the BS equipped with $ N $ transmit antennas and the $ S $ IRSs consisted of $ M $ elements. Moreover, We assume that every IRS allocated in the system is exactly the same. What's more, there is an occlusion between the BS and the range of multiple users, so we don’t need to consider the channel between the BS and users. The reflection matrix of IRS which is a diagonal matrix composed of the reflection coefficients of all elements is denoted as $\mathbf{\Phi}_{s}=\mathrm{diag}(\beta_1e^{j\theta _{1}},\beta_2e^{j\theta _{2}},\dots,\beta_Me^{j\theta _{M}} ) $.
In addition, we assume that $ \beta_m $, $ m\in (1,2,\cdots,M) $ is always equal to $1$ such that only the phase of the reflection coefficient $\theta _{1},\theta _{2},\dots,\theta _{M}\in [0,2\pi)$ can be adjusted. 

The complex transmitted signal from the BS is given by
\begin{equation}
	\mathbf{x}=\sum_{k=1}^{K}\mathbf{v}_{k}s_{k},
\end{equation}
where $ \mathbf{v}_{k} $ denotes beamforming vector from the BS to the $ k $-th user. The  transmitted symbols for $ k $-th user is denoted by $ s_{k} $, which satisfies the condition of $ \mathbb{E}\{s_{k}s_{k}^{H}\}=1 $. Then, the received signal $ y_k $ is expressed as\begin{equation}
	y_{k}=(\sum_{s=1}^{S}\mathbf{g}_{s,k}^{H}\mathbf{\Phi }_{s}\mathbf{H}_{s})\mathbf{x}+u_{k},
\end{equation}
where $\mathbf{H}_{s}\in \mathbb{C}^{M\times N} $ denotes the channel between the BS and the $s$-th IRS, $\mathbf{g}_{s,k}^{H}\in \mathbb{C}^{1\times M} $ denotes the channel between the $s$-th IRS and the $k$-th user and $ u_{k} $ is the received noise for $ k $-th user which follows the additive white Gaussian noise (AWGN) distribution of $ \mathcal{CN}\sim (0,\sigma^{2}_{k}) $.

In this paper, by employing the classic Saleh-Valenzula channel model \cite{el2014spatially}, we express $ \mathbf{H}_{s} $ and $ \mathbf{g}_{s,k}^{H} $ as
\begin{equation}
	\mathbf{H}_{s}=\sqrt{\frac{N M}{\varrho_\mathrm{BI} }} \sum_{l=0}^{L} \alpha_{l} \mathbf{a}_{\mathrm{IRS}}\left(\phi_{ l}^{r}, \varphi_{l}^{r}\right) \mathbf{a}_{\mathrm{BS}}\left(\phi_{ l}^{t}, \varphi_{l}^{t}\right)^{H},\label{h}
\end{equation}
and 
\begin{equation}
	\mathbf{g}_{s,k}^{H}=\sqrt{\frac{ M}{\varrho_\mathrm{IU} }} \sum_{l=0}^{L} \alpha_{l} \mathbf{a}_{\mathrm{IRS}}\left(\phi_{ l}^{r}, \varphi_{l}^{r}\right)^{H}\label{gh},
\end{equation}
where $ l = 0 $ represents the line-of-sight (LoS) path; $ \mathbf{H}_{s} $  and $ \mathbf{g}_{s,k}^{H} $ has $ L $ NLoS paths; $ \varrho_\mathrm{BI} $, $ \varrho_\mathrm{IU} $ denote the path-loss and $ \alpha_{l} $ is the complex gain of the $ l $-th path. Here, the azimuth and elevation angles at receiver are denoted by $ \phi_{ l}^{r}, \varphi_{l}^{r} $, and the azimuth and elevation angles at transmiter are denoted by $ \phi_{ l}^{t}, \varphi_{l}^{t} $. In \eqref{h} and \eqref{gh}, $ \mathbf{a}_{\mathrm{IRS}}\left(\phi, \varphi\right)$ and $ \mathbf{a}_{\mathrm{BS}}\left(\phi, \varphi\right) $ both satisfy this structure
\begin{equation}
	\begin{aligned}
	&\mathbf{a}(\phi, \varphi)= \frac{1}{\sqrt{RC}}\\
	&\times\left [e^{j\frac{2 \pi}{\lambda} d\bigl(0\sin (\phi) \sin (\varphi)\bigr)},\cdots, e^{j\frac{2 \pi}{\lambda} d\bigl((R-1)\sin (\phi) \sin (\varphi)\bigr)} \right ]^{T}\\
	& \otimes\left [ e^{j\frac{2 \pi}{\lambda} d\bigl(0\cos (\varphi)\bigr)},\cdots, e^{j\frac{2 \pi}{\lambda} d\bigl((C-1)\cos (\varphi)\bigr)} \right ]^{T}	,
	\end{aligned}
\end{equation}
where  $ R $ and $ C $ represent the number of antennas in the rows and columns of the antenna array respectively. $ \lambda $ is the wavelength and $ d $ is the distance between adjacent antenna elements which can be assumed to be $ \lambda/2 $.

Therefore, the achievable rate for $ k $-th user, $k=1, \dots, K$, is
\begin{equation}\label{Rk}
	\mathcal{R}_{k}=\log _{2}(1+r_{k}),
\end{equation}
where 
%$ r_{k} $ in (\ref{Rk}) is the Signal to Interference plus Noise Ratio (SINR) of $ k $-th user and can be expressed as
\begin{equation}
	r_{k}=\frac{|\mathbf{g}_{k}^{H}\mathbf{\Phi }\mathbf{H}\mathbf{v}_{k}|^{2}}{\sum_{j\neq k}^{K}|\mathbf{g}_{k}^{H}\mathbf{\Phi }\mathbf{H}\mathbf{v}_{j}|^{2}+\sigma _{k}^{2}},
	\label{eq_SINR}
\end{equation}
%where $ \mathbf{g}_{k}^{H} $, $ \mathbf{H} $ and $ \mathbf{\Phi } $ is:
with $\mathbf{g}_{k}^{H}=\left[ \mathbf{g}_{1,k}^{H},\mathbf{g}_{2,k}^{H},\dots,\mathbf{g}_{S,k}^{H}\right]$, $\mathbf{H}=\left[ \mathbf{H}_{1}^T, \mathbf{H}_{2}^T,\dots,\mathbf{H}_{S}^T\right] ^{T}$, and $\mathbf{\Phi }=\operatorname{diag}(\mathbf{\Phi }_{1},  \mathbf{\Phi }_{2}, \dots, \mathbf{\Phi }_{S})$.
% \begin{equation}
% 	\mathbf{\Phi }=\begin{bmatrix}
% 		\mathbf{\Phi }_{1} & \cdots &0 \\ 
% 		\vdots & \ddots &\vdots\\ 
% 		0 & \cdots  & \mathbf{\Phi }_{S}
% 	\end{bmatrix}.
% \end{equation}

We assume that all IRSs have the same type. Therefore, the weight sum-rate maximization optimization problem ($\mathrm{P1}$) for multi-user can be formulated as 
\begin{equation}\label{P1}
\begin{aligned}
	\left( \mathrm{P1}\right) :&\;\underset{\mathbf{V},\mathbf{\Phi} }{\mathrm{max}}\: f_{1}(\mathbf{V},\mathbf{\Phi} )=\sum_{k=1}^{K}\omega _{k}\mathcal{R}_{k}
	\\&\mathrm{s.t.}\; \mathrm{tr}(\mathbf{VV}^{H})\leq P,
	\\&\theta_{i}\in \mathcal{F},i=(1,2,\dots,SM),
\end{aligned}
\end{equation}
where $\mathbf{V}=[\mathbf{v}_{1},\mathbf{v}_{2},\cdots,\mathbf{v}_{K}] $ denotes the beamforming matrix; the weight $ \omega _{k} $ is the required service priority of the $ k $-th user and $ P $ is the maximum transmit power at BS; $\mathrm{tr}(\mathbf{VV}^{H})\leq P$ represents the BS transmit power constraint; $ \mathcal{F} $ denotes the feasible set of the reflection coefficient which has two different assumptions. One is a set of infinite resolution under ideal conditions, and the other is a set of finite resolution controlled by $ Q $-level quantization order. 
%The specific details will be explained in section \ref{E}.

\section{Manifold Optimization Based Algorithm}\label{3}
In this section, we propose a manifold optimization based algorithm which jointly optimize the beamforming matrix and reflection matrix using manifold optimization to find the optimal solution to problem ($\mathrm{P1}$).

\subsection{Reformulate the Problem}\label{A}
To deal with the sum-log problem, we decouple and divide the problem into three sub-problems \cite{free,shen2018fractional}. By introducing auxiliary variables $\bm{\gamma}=\left[\gamma_{1},\gamma_{2},\cdots,\gamma_{K} \right]  \in \mathbb{C}^{1\times K} $, the original problem  ($\mathrm{P1}$) is reformulated as
\begin{equation}
\begin{split}
\left( \mathrm{P2}\right): &\;\underset{\mathbf{V},\mathbf{\Phi},\mathbf{\bm{\gamma} }}{\mathrm{max}} \;f_{2}(\mathbf{V},\mathbf{\Phi},\mathbf{\bm{\gamma} })=\frac{1}{\ln 2}\sum_{k=1}^{K}\omega _{k}\ln(1+\mathbf{\gamma }_{k})\\&+\sum_{k=1}^{K}\bigl(-\omega _{k}\mathbf{\gamma }_{k}+\frac{\omega _{k}(1+\mathbf{\gamma }_{k})r_{k}}{1+r_{k}}\bigr)
\\& \mathrm{s.t.}\; \mathrm{tr}(\mathbf{VV}^{H})\leq P,
\\&\theta_{i}\in \mathcal{F},i=(1,2,\dots,SM).
\end{split}
\end{equation}

Then we alternatively optimize $ \mathbf{V}$ , $ \mathbf{\Phi} $  and $ \mathbf{\bm{\gamma} } $ jointly.
The key idea behind is to fix the other two variables and optimize the remained one until the convergence of the objective function is achieved.

\subsection{Fix $ \mathbf{V},\mathbf{\Phi} $ and  Optimize $ \mathbf{\bm{\gamma} } $}\label{B} 
During each iteration, the axuiliary variable $ \mathbf{\bm{\gamma} }$ is updated firstly according to $\mathbf{V}$ and $\mathbf{\Phi}$ at the previous iteration, via setting $\partial f_{2}(\mathbf{V},\mathbf{\Phi},\mathbf{\bm{\gamma} })/\partial \mathbf{\bm{\gamma} }  $ to zero. The updated $ \mathbf{\bm{\gamma} } $ is expressed as
\begin{equation}\label{gamma}
	{\mathbf{\gamma }}_{k}=r_{k}.
\end{equation}

Observe that $f_{2}$ is a concave differentiable function over $\bm{\gamma}$ when $ \mathbf{V}$ and $\mathbf{\Phi} $ is held fixed. Substituting \eqref{gamma} back in $f_{2}$ recovers the objective function in \eqref{P1} exactly. This is the reason why ($\mathrm{P2}$) is equal to ($\mathrm{P1}$). In addition, we notice that only one term of the objective function in  ($\mathrm{P2}$) is related to $ \mathbf{V}$  and $\mathbf{\Phi} $. So when optimizing $ \mathbf{V}$  and $\mathbf{\Phi} $, the objective function can be further simplified as
\begin{equation}
	\begin{aligned}
		\left( \mathrm{P3}\right):&\; 	\underset{\mathbf{V},\mathbf{\Phi }}{\mathrm{max}}\; f_{3}(\mathbf{V},\mathbf{\Phi })\\&=\sum_{k=1}^{K}\frac{\omega _{k}(1+\gamma _{k})r_{k}}{1+r_{k}}
		\\&=\sum_{k=1}^{K}\frac{\omega _{k}(1+\gamma _{k})|(\mathbf{g}_{k}^{H}\mathbf{\Phi }\mathbf{H})\mathbf{v}_{k}|^{2}}{\sum_{j=1}^{K}|(\mathbf{g}_{k}^{H}\mathbf{\Phi }\mathbf{H})\mathbf{v}_{j}|^{2}+\sigma _{k}^{2}}
		\\& \mathrm{s.t.}\; \mathrm{tr}(\mathbf{VV}^{H})\leq P,
		\\&\theta_{i}\in \mathcal{F},i=(1,2,\dots,SM).
	\end{aligned}
\end{equation}

\subsection{Fix  $ \mathbf{\bm{\gamma} } $ , $\mathbf{\Phi} $ and Optimize $ \mathbf{V} $}\label{C}
Given fixed $\mathbf{\Phi} $, ($\mathrm{P3}$)  is equivalent to
\begin{equation}
		\begin{aligned}
			\left( \mathrm{P3'}\right):&\;	\underset{\mathbf{V}}{\mathrm{max}}\; f_{3}(\mathbf{V})
			\\&=\sum_{k=1}^{K}\frac{\omega _{k}(1+\gamma _{k})|(\mathbf{g}_{k}^{H}\mathbf{\Phi }\mathbf{H})\mathbf{v}_{k}|^{2}}{\sum_{j=1}^{K}|(\mathbf{g}_{k}^{H}\mathbf{\Phi }\mathbf{H})\mathbf{v}_{j}|^{2}+\sigma _{k}^{2}}
			\\& \mathrm{s.t.}\; \mathrm{tr}(\mathbf{VV}^{H})\leq P,
		\end{aligned}
\end{equation}

Without loss of generality, we normalize the power to get $ \mathrm{tr}(\mathbf{VV}^{H})\leq 1 $ and define $ \mathbf{\hat{V}}=[\mathbf{\hat{v}}_{1},\mathbf{\hat{v}}_{2},\cdots,\mathbf{\hat{v}}_{K}]$ satisfying $ \mathrm{tr}(\mathbf{\hat{V}\hat{V}}^{H})=\mathrm{tr}(\mathbf{VV}^{H})+||\mathbf{\varkappa}||_2^2= 1$, where $ \mathbf{\hat{v}}_{k}=[\mathbf{v}_{k}^T,\varkappa_k]^T $ and $ \mathbf{\varkappa}=[\varkappa_1,\varkappa_2,\cdots,\varkappa_K] $ is an auxiliary vector to convert the unequal constraint to be an equal one. 
As the Frobenius norm is equal to one, we then define a $ (N+1)K-1 $ dimensional complex sphere manifold $ \mathcal{M}_{1}=\left\lbrace  \mathbf{\hat{V}}\in \mathbb{C}^{(N+1)\times K}\big| \mathrm{tr}(\mathbf{\hat{V}\hat{V}}^{H})= 1 \right\rbrace $, such that  ($\mathrm{P3'}$) is equivalent to the unconstrained optimization problem on $\mathcal{M}_{1} $, that is
\begin{equation}
\begin{aligned}
\left( \mathrm{P3''}\right):&\;	\underset{\mathbf{\hat{V}}\in \mathcal{M}_{1}}{\mathrm{max}}\; f_{3}(\mathbf{\hat{V}})
\\&=\sum_{k=1}^{K}\frac{\omega _{k}(1+\gamma _{k})|(\mathbf{g}_{k}^{H}\mathbf{\Phi }\mathbf{\hat{H}})\mathbf{\hat{v}}_{k}|^{2}}{\sum_{j=1}^{K}|(\mathbf{g}_{k}^{H}\mathbf{\Phi }\mathbf{\hat{H}})\mathbf{\hat{v}}_{j}|^{2}+\sigma _{k}^{2}},
\end{aligned}
\end{equation}
where $ \mathbf{\hat{H}}=\sqrt{P}[\mathbf{H},\mathbf{0}] $. 

Now let's review the geometric conjugate gradient(GCG) algorithm \cite{al} 
for this kind of manifold optimization problem, as given in \textbf{Algorithm 1}.
For the conjugate gradient algorithm in Euclidean space, the next iteration point can be determined by the step size and search direction. But if the selectable variables define a Riemannian submanifold, it is very likely that the next iteration point isn't on the manifold. Because of this, we need to make different retractions for different manifolds to ensure that the next iteration point is still on the manifold. In \textbf{Algorithm 1}, retraction is denoted as $ \mathcal{R}$. For example, the retraction of $ \alpha _{t}\eta _{t} $ ($ \alpha _{t} $ denotes  Armijo step size) at point $ \mathbf{x}_{t}   $ onto the manifold $ \mathcal{M} $ can be expressed as $ \mathcal{R}_{\mathbf{x}_{t}}(\alpha _{t}\eta _{t}) $. Moreover, if the tangent plane at point $ \mathbf{x}_{t} $ is defined as $ T\mathcal{M}_{\mathbf{x}_{t}} $, then we know that $\mathrm{Rgrad}\phi(\mathbf{x}_{t+1})$ means the Riemannian gradient at $ \mathbf{x}_{t+1}   $ belongs to $ T\mathcal{M}_{\mathbf{x}_{t+1}} $  while $\eta _{t}$  means the search direction at $ \mathbf{x}_{t} $ belongs to $ T\mathcal{M}_{\mathbf{x}_{t}} $. Since $ \mathrm{Rgrad}\phi(\mathbf{x}_{t+1}) $ and $ \eta _{t} $ are not in the same tangent plane, they cannot be summed directly. To this end, we need to transport $ \eta _{t} $ to $T\mathcal{M}_{\mathbf{x}_{t+1}} $. In \textbf{Algorithm 1}, the operator
  of vector transport is denoted as $ \mathcal{T} $.  For example, the vector transport of $ \eta _{t} $ to $T\mathcal{M}_{\mathbf{x}_{t+1}} $ by combining $ \mathcal{R}_{\mathbf{x}_{t}}(\alpha _{t}\eta _{t}) $  can be expressed as $\mathcal{T}_{\alpha _{t}\eta _{t}}(\eta _{t}) $. More \textbf{Algorithm 1} details and convergence analysis can be obtained in \cite{al}.%The core concepts is geometrically illustrated in Fig.\ref{g}. Among them, the red line represents retraction and the blue line represents vector transport.

As the GCG algorithm make use of first-order information of the cost function, the key step is to calculate the Riemannian gradient of the objective function at the current point. 
By calculating the Euclidean gradient of the cost function and projecting the Euclidean gradient onto tangent space \cite{boumal2020introduction}, we have the Riemannian gradient of cost function. % defined on a Riemannian submanifold of a Euclidean space can be calculated based on the projection of the Euclidean gradient onto tangent space \cite{boumal2020introduction}, thus we only need to calculate the Euclidean gradient of the cost function.
  
  \begin{algorithm}[ht] 
  	\caption{Geometric Conjugate Gradient Algorithm} 
  	\begin{algorithmic}[1] 
  		\Require Retraction $ \mathcal{R} $ on $ \mathcal{M} $ and vector transport $ \mathcal{T} $ on $ \mathcal{M} $, the cost function $ \phi  $;
  		\State Initialize  point $ \mathbf{x}_{0}\in \mathcal{M} $;
  		\State Initialize search direction $\eta _{0}=\mathrm{Rgrad}\phi(\mathbf{x}_{0}) $;
  		\State Iteration counter $ t=0 $;
  		\Repeat
  		\State Compute $\alpha_{t} > 0 $ according to ALS Algorithm in \cite{al};
  		\State Update point $ \mathbf{x}_{t+1}=\mathcal{R}_{\mathbf{x}_{t}}(\alpha _{t}\eta _{t}) $;
  		\State Compute correction parameter $ \beta _{t+1}=\frac{\left \langle \mathrm{Rgrad}\phi(\mathbf{x}_{t+1}), \mathrm{Rgrad}\phi(\mathbf{x}_{t+1}) \right \rangle}{\left \langle \mathrm{Rgrad}\phi(\mathbf{x}_{t}),\mathrm{Rgrad}\phi(\mathbf{x}_{t}) \right \rangle} $;
  		\State Update search direction $ \mathbf{\eta }_{t+1}=\mathrm{Rgrad}\phi(\mathbf{x}_{t+1})+\beta _{t+1}\mathcal{T}_{\alpha _{t}\eta _{t}}(\eta _{t}) $;
  		\State $t=t+1 $;
  		\Until Convergence;
  		\Ensure Converged point $\mathbf{x}_{t}  $.
  	\end{algorithmic} 
  \end{algorithm}

%   \begin{figure}[h]
%   	\centerline{\includegraphics[width=0.3\textwidth]{Retraction_and_vector_transport.eps}}
%   	\caption{Geometric interpretation of retraction and vector transport.}
%   	\label{g}
%   \end{figure}

For our problem, we can't use GCG algorithm directly because the variable to be optimized is $ \hat{\mathbf{v}}_{1},\cdots,\hat{\mathbf{v}}_{K} $, which is a certain column of $ \hat{\mathbf{V}} $, but the constraint is about $ \hat{\mathbf{V}} $. Therefore, we construct an index matrix which is a $ k $-order identity matrix $\mathbf{E}_{k} $ so that any column of $ \hat{\mathbf{V}} $ can be expressed by $ \hat{\mathbf{V}} $ and the index matrix. If the $ i $-th column of $\mathbf{E}_{k} $ is expressed as $\mathbf{E}_{ki} $, then ($ \mathrm{P3''} $) is rewritten as
\begin{equation}\label{P4}
\begin{aligned}
	\left( \mathrm{P4}\right):&\underset{\hat{\mathbf{V}}\in\mathcal{M}_{1}}{\mathrm{max}} f_{4}(\hat{\mathbf{V}})=\\&\sum_{k=1}^{K}\frac{\tilde{\gamma} _{k}|(\mathbf{g}_{k}^{H}\mathbf{\Phi }\hat{\mathbf{H}})\hat{\mathbf{V}}\mathbf{E}_{kk}|^{2}}{\sum_{j=1}^{K}|(\mathbf{g}_{k}^{H}\mathbf{\Phi }\hat{\mathbf{H}})\hat{\mathbf{V}}\mathbf{E}_{kj}|^{2}+\sigma _{k}^{2}},
\end{aligned}
\end{equation}
where $\tilde{\gamma }_{k}$ is equivalent to $ \omega _{k}(1+\gamma _{k}) $.

%Since the expression of the Euclidean gradient of the objective function for $ \hat{\mathbf{V}} $ in ($ \mathrm{P4} $) is too complicated, 
Define $ \tilde{\mathbf{h}}_{k}^{H}=\mathbf{g}_{k}^{H}\mathbf{\Phi }\hat{\mathbf{H}} $, then the Euclidean gradient is simplified to 
\begin{equation}\label{EV}
\begin{split}
&\mathrm{Egrad}f_{4}(\hat{\mathbf{V}})
=\sum_{k=1}^{K}\tilde{\gamma }_{k}\cdot \left(\frac{2\tilde{\mathbf{h}}_{k}^{H}\hat{\mathbf{V}}\mathbf{E}_{kk}\tilde{\mathbf{h}}_{k}\mathbf{E}_{kk}^{H}}{\sum_{j=1}^{K}|\tilde{\mathbf{h}}_{k}^{H}\hat{\mathbf{V}}\mathbf{E}_{kj}|^{2}+\sigma _{k}^{2}}\right.\\&\left.-2\sum_{i=1}^{K}\frac{|\tilde{\mathbf{h}}_{k}^{H}\hat{\mathbf{V}}\mathbf{E}_{kk}|^{2}\tilde{\mathbf{h}}_{k}^{H}\hat{\mathbf{V}}\mathbf{E}_{ki}\tilde{\mathbf{h}}_{k}\mathbf{E}_{ki}^{H}}{(\sum_{j=1}^{K}|\tilde{\mathbf{h}}_{k}^{H}\hat{\mathbf{V}}\mathbf{E}_{kj}|^{2}+\sigma _{k}^{2})^{2}}\right).
\end{split}
\end{equation}

The projection operator of the tangent plane at point $ \hat{\mathbf{V}} $ on $ \mathcal{M}_{1} $ is defined as
\begin{equation}\label{PV}
\mathbf{P}_{ \hat{\mathbf{V}}}\left( \mathbf{\Psi}\right) = \mathbf{\Psi} - \mathrm{tr}(\hat{\mathbf{V}}^H \mathbf{\Psi}) \hat{\mathbf{V}}, 
\end{equation}

where $ \mathbf{\Psi} $ represents a matrix in the ambient space.

Then the Riemann gradient can be obtained according to (\ref{EV}) and (\ref{PV}), expressed as
\begin{equation}
\mathrm{Rgrad}f_{4}(\hat{\mathbf{V}})=\mathbf{P}_{ \hat{\mathbf{V}}}\left( \mathrm{Egrad}f_{4}(\hat{\mathbf{V}})\right).
\end{equation}

Moreover, the retraction and vector transport on complex sphere are
\begin{equation}
\mathcal{R}_{\mathbf{\hat{V}}_{t}}(\alpha _{t}\eta _{t})_{\mathcal{SP}} =\dfrac{\mathbf{\hat{V}}_{t}+\alpha _{t}\eta _{t}}{||\mathbf{\hat{V}}_{t}+\alpha _{t}\eta _{t}||},
\end{equation}
and
\begin{equation}
\mathcal{T}_{\alpha _{t}\eta _{t}}(\eta _{t})_{\mathcal{SP}}=\mathbf{P}_{ \mathcal{R}_{\mathbf{\hat{V}}_{t}}(\alpha _{t}\eta _{t})}\left( \eta _{t}\right).
\end{equation}

The superscript $ (\cdot)_\mathcal{SP} $ means that the retraction and vector transport is on the complex sphere manifold. 

Finally, we can optimize $ \mathbf{\hat{V}} $ by applying \textbf{Algorithm 1} and $\mathbf{V} $ can be obtained via $ \mathbf{V}=\mathbf{\hat{V}}(1:N,K) $.
\subsection{Fix  $ \mathbf{\bm{\gamma} } $ , $ \mathbf{V} $ and Optimize $\mathbf{\Phi} $}\label{D}

In this subsection, we adopt a similar manifold optimization approach to the previous subsection. Define $\mathbf{u}^{*}=\mathrm{vec}(\mathbf{\Phi}) $, we have $\mathbf{u}=\left ( e^{j\theta _{1}},e^{j\theta _{2}},...,e^{j\theta _{SM}}\right ) ^{H}$ satisfies $ (\mathbf{u}\mathbf{u}^H)_{ii} = 1$, $i = 1:SM$. Then $\mathbf{u} $ forms a complex oblique manifold named $ \mathcal{M}_{2} $. Naturally, we can derive $\mathbf{u}^{H}\mathrm{diag}(\mathbf{g}_{k}^{H})\mathbf{H}=\mathbf{g}_{k}^{H}\mathbf{\Phi }\mathbf{H} $ and rewrite ($ \mathrm{P3} $) as
\begin{equation}\label{P5}
\begin{aligned}
\left( \mathrm{P5}\right):&\;\underset{\mathbf{u}\in \mathcal{M}_{2}}{\mathrm{max}} \;f_{5}(\mathbf{u})\\&=\sum_{k=1}^{K}\frac{\tilde{\gamma }_{k}|\bigl(\mathbf{u}^{H}\mathrm{diag}(\mathbf{g}_{k}^{H})\mathbf{H}\bigr )\mathbf{V}_{k}|^{2}}{\sum_{j=1}^{K}|\bigl(\mathbf{u}^{H}\mathrm{diag}(\mathbf{g}_{k}^{H})\mathbf{H}\bigr )\mathbf{V}_{j}|^{2}+\sigma _{k}^{2}}
\\&\mathrm{s.t.}\;	\theta_{i}\in \mathcal{F},i=(1,2,\dots,SM).
\end{aligned}
\end{equation}

The projection operator of the tangent plane at point $ \mathbf{u} $ on $ \mathcal{M}_{2} $ \cite{hu2020brief} can be characterized by
\begin{equation}
\mathbf{P}_{ \mathbf{u}}\left( \bm{\psi}\right) = \bm{\psi} - \operatorname{ddiag}(\bm{\psi}\mathbf{u}^H)\mathbf{u}.
\end{equation}

where $ \bm{\psi} $ represents a vector in the ambient space.

Then we calculate the Euclidean gradient and the Riemann gradient of $ f_{5}(\mathbf{u}) $ according to (\ref{equ}) at the top of next page and 
\begin{figure*}
\begin{equation}\label{equ}
\begin{split}
&\mathrm{Egrad}f_{5}(\mathbf{u })=\sum_{k=1}^{K}\tilde{\gamma }_{k} \left(\frac{2\mathbf{v}_{k}^{H}(\mathbf{H}^{H}\mathrm{diag}^{H}(\mathbf{g}_{k}^{H})\mathbf{u})\mathrm{diag}(\mathbf{g}_{k}^{H})\mathbf{H}\mathbf{v}_{k}}{\sum_{j=1}^{K}|(\mathbf{u}^{H}\mathrm{diag}(\mathbf{g}_{k}^{H})\mathbf{H})\mathbf{v}_{j}|^{2}+\sigma _{k}^{2}}-\sum_{i=1}^{K}\frac{2|(\mathbf{u}^{H}\mathrm{diag}(\mathbf{g}_{k}^{H})\mathbf{H})\mathbf{v}_{k}|^{2}\mathbf{v}_{i}^H(\mathbf{H}^{H}\mathrm{diag}^{H}(\mathbf{g}_{k}^{H})\mathbf{u})\mathrm{diag}(\mathbf{g}_{k}^{H})\mathbf{H}\mathbf{v}_{i}}{(\sum_{j=1}^{K}|(\mathbf{u}^{H}\mathrm{diag}(\mathbf{g}_{k}^{H})\mathbf{H})\mathbf{v}_{j}|^{2}+\sigma _{k}^{2})^{2}}\right).
\end{split}
\end{equation}
\end{figure*}

\begin{equation}
\mathrm{Rgrad}f_{5}(\mathbf{u})=\mathbf{P}_{ \mathbf{u}}\left( \mathrm{Egrad}f_{5}(\mathbf{u})\right),
\label{eq_ReinmanGrandient}
\end{equation}
respectively. 
Finally, the retraction and vector transport on complex oblique manifold are given by
\begin{equation}
\begin{aligned}
&\mathcal{R}_{\mathbf{u}_{t}}(\alpha _{t}\eta _{t})_{\mathcal{OB}}=\\
&\mathrm{ddiag}\left(\left( \mathbf{u}_{t}+\alpha _{t}\eta _{t}\right)\left( \mathbf{u}_{t}+\alpha _{t}\eta _{t}\right)^H\right)^{-1/2}\left( \mathbf{u}_{t}+\alpha _{t}\eta _{t}\right),
\end{aligned}
\end{equation}
and 
\begin{equation}
\mathcal{T}_{\alpha _{t}\eta _{t}}(\eta _{t})_{\mathcal{OB}}=\mathbf{P}_{ \mathcal{R}_{\mathbf{u}_{t}}(\alpha _{t}\eta _{t})}\left( \eta _{t}\right),
\end{equation}
%where the function of \textcolor{red}{$ \operatorname{normalize} $} is to make every element of the vector have a norm $1$.
where $ (\cdot)_\mathcal{OB} $ means the retraction and vector transport on the complex oblique manifold.

\begin{algorithm}[t] 
	\caption{Double Manifold Alternating Optimization Algorithm} 
	\begin{algorithmic}[1] 
		\Require Channel State Information about $ \mathbf{H} $ and $ \mathbf{g}_{1}$,$ \mathbf{g}_{2} $,$ \cdots $,$ \mathbf{g}_{K} $; user service priority  $ \omega _{k} $.
		\State Initialize $ \mathbf{\hat{V}}$ and  $ \mathbf{u} $;
		\Repeat
		\State Update $ \mathbf{\bm{\gamma}} $ by (\ref{gamma});
		\State Update $ \mathbf{\hat{V}}$ to solve (\ref{P4}) using \textbf{Algorithm 1};
		\State Update $ \mathbf{u}$ to solve (\ref{P5}) using \textbf{Algorithm 1};
		\Until Convergence;
		\Ensure Optimized beamforming vector $ \mathbf{v}_{1}$,$ \mathbf{v}_{2}$,$ \cdots $,$ \mathbf{v}_{K}$ for all users; Optimized reflection matrix $ \mathbf{\mathbf{\Phi}} $; Local optimal solution of $ f_{1}(\mathbf{V},\mathbf{\Phi} ) $.
	\end{algorithmic} 
\end{algorithm}

The optimization process are summarized in \textbf{Algorithm 2}, namely double manifold alternating optimization algorithm (DMAO). Notice that the monotonic property of \textbf{Algorithm 2} along with the fact that the objective function is upper-bounded, is sufficient to prove the convergence. Moreover, the computational complexity of \textbf{Algorithm 2} is mainly due to the \textbf{Algorithm 1} iterations. Specifically, the calculation of $ \mathbf{\hat{V}} $ requires $ \mathcal{O}\left( KSM\cdot\max\{SM, N\} +NK^3\right)  $ operations, and solving (\ref{P5}) requires $ \mathcal{O}\left( S^2M^2NK+SMNK^2+S^3M^3\right)  $ operations per-iteration of \textbf{Algorithm 1}.
\subsection{Non-ideal IRS Case}\label{E}
For non-ideal IRS, due to hardware limitations, it is impossible to set the reflection coefficient to an arbitrary value, that is to say, it is impossible to meet the requirement of infinite resolution of the reflection angle. Therefore, in this subsection, the reflection angle is quantified so that it can only be 
%optimized to certain specific angles. The quantitative idea we adopted is to compare the angle distance between the optimized angle and each specific angle, and 
selected from the feasible angle set with the shortest distance from the optimized value. Quantification is realized by
\begin{equation}
\bar{\theta }_{i}=\underset{\theta ^{'}\in \mathcal{F}}{\mathrm{argmin}}|\theta _{i}-\theta^{'}|,\mathcal{F}=\{0,\frac{2\pi }{Q},\cdots \frac{2\pi (Q-1)}{Q}\},
\end{equation}
where $\bar{\theta }_{i} $, $\theta _{i} $ and $\theta^{'} $ is the reflection angle of the final output, infinite resolution reflection angle obtained by \textbf{Algorithm 2} and feasible angle, respectively. $ \mathcal{F} $ represents the set of feasible angles, which is a set of finite resolution controlled by $ Q $-level quantization order as mentioned in \ref{2}.

\section{Simulation Results}\label{4}
%In this section, we provided many simulation results under different conditions to evaluate the performance of the proposed algorithm.

Under the framework of the multiuser MISO system in this paper, we assume that all users have single antenna. For BS and all IRSs, the number of antennas in row are fixed and only the number of column antennas can be adjusted. We suppose that the deployed IRSs have the same type. 
The variance of AWGN for each user is set as $\sigma^{2}_k=-80\mathrm{dBm} $. 
The maximum transmit power is set as $ P=1\mathrm{W} $. For simplicity, the weight $ \omega_k $ are set to be $1$.

$ \mathbf{H}_s $ and $ \mathbf{g}^H_{s,k} $, $ s=(1,\cdots,S) $ and $ k=(1,\cdots,K) $, a total of $ S(K+1) $ channels, all have $  L=3 $ scattering paths with $ \alpha_l\sim\mathcal{CN} (0,0.4) $. In addition, $ \alpha_0 $ satisfies $ \alpha_0\sim\mathcal{CN} (0,2) $. All azimuth and elevation angles are uniformly distributed in $ [0,2\pi) $. The path-loss including $ \varrho_\mathrm{BI} $ and $ \varrho_\mathrm{IU} $ is given by
\begin{equation}
\varrho =(\frac{4\pi f_{\mathrm{carrier}}D}{c})^{2},
\end{equation}
where $ c $ is the speed of light and $ D $ is the distance between two points. $ f_{\mathrm{carrier}} $ is the carrier frequency and is set to 3GHz. We use the Monte Carlo method to simulate the channel realizations under different conditions 500 times in this paper. 

We assume a simulation scenario, where the coordinate of the base station is (0, 0), and the coordinates of the two IRSs are (10,24) and (24,10), respectively. Users are uniformly distributed in a circular area with a radius of 2 meters, and the center coordinate is (20, 0).

Next, we compare our proposed algorithms with the following four benchmark schemes.

1) \textbf{Benchmark scheme with random reflection matrix}: The reflection matrix is set randomly. In addition, the beamforming matrix is updated by solving problem $ \left( \mathrm{P4}\right)  $.

2) \textbf{Alternating optimization with maximum ratio transmission (MRT)}: 
The beamforming matrix is set based on the MRT principle as
% \begin{equation}
$	\mathbf{V}=\sqrt{\frac{P}{\left\| \mathbf{H}_{MRT} \right\| ^2_F } }\mathbf{H}_{MRT}^H$,
% \end{equation}
where $ \mathbf{H}_{MRT} $ is equal to $ \mathbf{g}_{k}^{H}\mathbf{\Phi }\mathbf{H} $, while the reflection matrix is updated by solving problem $ \left( \mathrm{P5}\right)  $. We optimize the  beamforming matrix and the reflection matrix via an alternating manner.

3) \textbf{Alternating optimization with zero forcing (ZF)}: The beamforming matrix is set based on the ZF principle as
$\mathbf{V}=\sqrt{\frac{P}{\mathrm{tr}\left(\left( \mathbf{H}_{ZF}\mathbf{H}_{ZF}^H \right) ^{-1}\right)  } }\mathbf{H}_{ZF}^H\left( \mathbf{H}_{ZF}\mathbf{H}_{ZF}^H \right) ^{-1}$,
where $ \mathbf{H}_{ZF} $ is equal to $ \mathbf{g}_{k}^{H}\mathbf{\Phi }\mathbf{H} $. On the other hand, the reflection matrix are obtained by solving problem $ \left( \mathrm{P5}\right)  $. We also use an alternating manner to optimize the  beamforming matrix and the reflection matrix.
 
4) \textbf{Algorithm proposed in \cite{scene}}:
%\textcolor{red}{add some explaining}
The beamforming matrix and the reflection matrix are optimized alternately with closed-form expressions, and related auxiliary variables are resolved using Lagrangian multiplier method, bisection search method and CVX toolbox \cite{cvx}.

\begin{figure}[htb]
	\centerline{\includegraphics[width=0.31\textwidth]{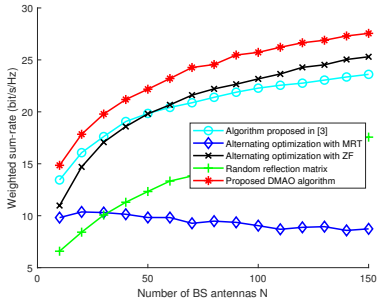}}
	\caption{Weighted sum-rate against the number of BS antennas.}
	\label{n}
\end{figure}

%To analyze the effect of BS antennas on the simulation scenario, we show 
Let each IRS has $ M=20 $ elements and the number of users is $ K=4 $, the weighted sum-rate against the BS antennas is shown in Fig. \ref{n}. It can be seen that as the number of BS antennas increases, the weighted sum-rate increases monotonically. The reason is that increasing the number of BS antennas can improve the channel conditions, which has been explained in related theories of MIMO. But we also see that the upward trend is getting slower and slower, because the key to determining the weighted sum-rate lies in the transmit power, but the maximum transmit power is set to $ 1\mathrm{W} $ and has not changed. What's more, for any BS antennas $N$, DMAO algorithm we proposed has an overwhelming advantage over other benchmark schemes. It is also observed that the performance achieved by ZF is better than the performance achieved by the algorithm proposed in \cite{scene} when $ N $ is large.

\begin{figure}[htb]
	\centerline{\includegraphics[width=0.31\textwidth]{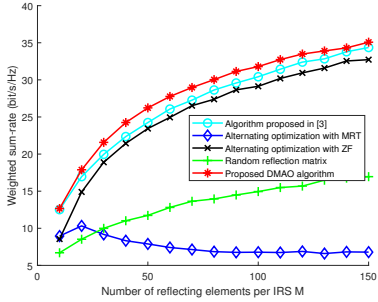}}
	\caption{Weighted sum-rate against the number of reflecting elements per IRS.}
	\label{m}
\end{figure}

In Fig. \ref{m}, we plot the weighted sum-rate against the number of reflecting elements where the number of BS antennas is fixed at $ N=20 $ and $ K=4 $. Note that as the number of elements of IRS grows, the sum-rate also increases. But with the growth of $ M $, the weighted sum-rate grows slower and slower. On the one hand, the greater the number of reflecting elements, the more incident signal that can be reused. On the other hand, the reflecting elements of IRS is passive elements which can only reuse the incident signal, so there is an upper bound on the improvement of weighted sum-rate. DMAO is also better than other benchmark schemes in this setting. 

\begin{figure}[htb]
	\centerline{\includegraphics[width=0.31\textwidth]{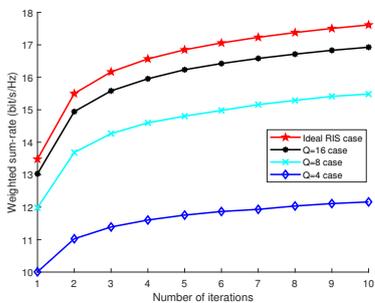}}
	\caption{Weighted sum-rate against quantization order.}
	\label{f}
\end{figure}

At last, we simulate scenarios with non-ideal IRS. The parameter is set to $ N,M=20 $ and $ K=4 $. Fig. \ref{f} shows the performance under different quantifications. As the quantization order $ Q $ increases, the weight sum-rate increases. When $ Q $ approaches infinity, the performance will coincide with the ideal performance. Generally, $ Q $ needs to satisfy $ Q=2^B $, which means that the element is controlled by $ B $ bits. 

\section{Conclusion}\label{5}
In this paper, we proposed the DMAO Algorithm to solve the weighted sum-rate maximization problem in multiuser system by jointly optimizing the beamforming matrix of the BS and reflection coefficient of each IRS element. The utilization of the manifold geometry are proved to be effective. Moreover, we also give an optimization method for the reflection coefficient in the case of non-ideal IRS. The simulation results have demonstrated that the proposed algorithm outperforms benchmark approaches and the system performance has a upper bound with the increase of the number of BS antennas and reflecting elements. It is also shown that the weight sum-rate will rise by increasing the quantization order.

\bibliographystyle{IEEEtran}
\bibliography{ref}

\begin{thebibliography}{10}
\providecommand{\url}[1]{#1}
\csname url@samestyle\endcsname
\providecommand{\newblock}{\relax}
\providecommand{\bibinfo}[2]{#2}
\providecommand{\BIBentrySTDinterwordspacing}{\spaceskip=0pt\relax}
\providecommand{\BIBentryALTinterwordstretchfactor}{4}
\providecommand{\BIBentryALTinterwordspacing}{\spaceskip=\fontdimen2\font plus
\BIBentryALTinterwordstretchfactor\fontdimen3\font minus
  \fontdimen4\font\relax}
\providecommand{\BIBforeignlanguage}[2]{{%
\expandafter\ifx\csname l@#1\endcsname\relax
\typeout{** WARNING: IEEEtran.bst: No hyphenation pattern has been}%
\typeout{** loaded for the language `#1'. Using the pattern for}%
\typeout{** the default language instead.}%
\else
\language=\csname l@#1\endcsname
\fi
#2}}
\providecommand{\BIBdecl}{\relax}
\BIBdecl

\bibitem{5G}
F.~Boccardi, R.~W. Heath, A.~Lozano, T.~L. Marzetta, and P.~Popovski, ``Five
  disruptive technology directions for 5{G},'' \emph{IEEE communications
  magazine}, vol.~52, no.~2, pp. 74--80, 2014.

\bibitem{myth}
E.~Bj{\"o}rnson, {\"O}.~{\"O}zdogan, and E.~G. Larsson, ``Reconfigurable
  {I}ntelligent {S}urfaces: Three myths and two critical questions,''
  \emph{IEEE Communications Magazine}, vol.~58, no.~12, pp. 90--96, 2020.

\bibitem{scene}
Y.~Cao and T.~Lv, ``Intelligent {R}eflecting {S}urface aided multi-user
  millimeter-wave communications for coverage enhancement,'' \emph{arXiv
  preprint arXiv:1910.02398}, 2019.

\bibitem{towards}
Q.~Wu and R.~Zhang, ``Towards smart and reconfigurable environment: Intelligent
  {R}eflecting {S}urface aided wireless network,'' \emph{IEEE Communications
  Magazine}, vol.~58, no.~1, pp. 106--112, 2019.

\bibitem{MIMO1}
E.~Bj{\"o}rnson and L.~Sanguinetti, ``Power scaling laws and near-field
  behaviors of massive {MIMO} and {I}ntelligent {R}eflecting {S}urfaces,''
  \emph{IEEE Open Journal of the Communications Society}, vol.~1, pp.
  1306--1324, 2020.

\bibitem{li2019joint}
Y.~Li, M.~Jiang, Q.~Zhang, and J.~Qin, ``Joint beamforming design in
  multi-cluster {MISO} {NOMA} {I}ntelligent {R}eflecting {S}urface-aided
  downlink communication networks,'' \emph{arXiv preprint arXiv:1909.06972},
  2019.

\bibitem{wu2020joint}
Q.~Wu and R.~Zhang, ``Joint active and passive beamforming optimization for
  {I}ntelligent {R}eflecting {S}urface assisted {SWIPT} under {QoS}
  constraints,'' \emph{IEEE Journal on Selected Areas in Communications}, 2020.

\bibitem{cui2019secure}
M.~Cui, G.~Zhang, and R.~Zhang, ``Secure wireless communication via
  {I}ntelligent {R}eflecting {S}urface,'' \emph{IEEE Wireless Communications
  Letters}, vol.~8, no.~5, pp. 1410--1414, 2019.

\bibitem{xu2019resource}
D.~Xu, X.~Yu, Y.~Sun, D.~W.~K. Ng, and R.~Schober, ``Resource allocation for
  secure {IRS}-assisted multiuser {MISO} systems,'' in \emph{2019 IEEE Globecom
  Workshops (GC Wkshps)}.\hskip 1em plus 0.5em minus 0.4em\relax IEEE, 2019,
  pp. 1--6.

\bibitem{yu2019miso}
X.~Yu, D.~Xu, and R.~Schober, ``{MISO} wireless communication systems via
  {I}ntelligent {R}eflecting {S}urfaces,'' in \emph{2019 IEEE/CIC International
  Conference on Communications in China (ICCC)}.\hskip 1em plus 0.5em minus
  0.4em\relax IEEE, 2019, pp. 735--740.

\bibitem{chen2019sum}
W.~Chen, X.~Ma, Z.~Li, and N.~Kuang, ``Sum-rate maximization for {I}ntelligent
  {R}eflecting {S}urface based terahertz communication systems,'' in \emph{2019
  IEEE/CIC International Conference on Communications Workshops in China (ICCC
  Workshops)}.\hskip 1em plus 0.5em minus 0.4em\relax IEEE, 2019, pp. 153--157.

\bibitem{guo2019weighted}
H.~Guo, Y.-C. Liang, J.~Chen, and E.~G. Larsson, ``Weighted sum-rate
  optimization for {I}ntelligent {R}eflecting {S}urface enhanced wireless
  networks,'' \emph{arXiv preprint arXiv:1905.07920}, 2019.

\bibitem{li2020manifold}
J.~Li, G.~Liao, Y.~Huang, and A.~Nehorai, ``Manifold optimization for joint
  design of {MIMO}-{STAP} radars,'' \emph{IEEE Signal Processing Letters},
  2020.

\bibitem{lin2020channel}
T.~Lin, X.~Yu, Y.~Zhu, and R.~Schober, ``Channel estimation for {I}ntelligent
  {R}eflecting {S}urface-assisted millimeter wave {MIMO} systems,'' \emph{arXiv
  preprint arXiv:2005.04720}, 2020.

\bibitem{hong2020semi}
X.~Hong, J.~Gao, and S.~Chen, ``Semi-blind joint channel estimation and data
  detection on sphere manifold for {MIMO} with high-order {QAM} signaling,''
  \emph{Journal of the Franklin Institute}, 2020.

\bibitem{douik2019manifold}
A.~Douik and B.~Hassibi, ``Manifold optimization over the set of doubly
  stochastic matrices: A second-order geometry,'' \emph{IEEE Transactions on
  Signal Processing}, vol.~67, no.~22, pp. 5761--5774, 2019.

\bibitem{el2014spatially}
O.~El~Ayach, S.~Rajagopal, S.~Abu-Surra, Z.~Pi, and R.~W. Heath, ``Spatially
  sparse precoding in millimeter wave {MIMO} systems,'' \emph{IEEE transactions
  on wireless communications}, vol.~13, no.~3, pp. 1499--1513, 2014.

\bibitem{free}
Z.~Zhang and L.~Dai, ``A joint precoding framework for wideband
  {R}econfigurable {I}ntelligent {S}urface-aided cell-free network,''
  \emph{arXiv preprint arXiv:2002.03744}, 2020.

\bibitem{shen2018fractional}
K.~Shen and W.~Yu, ``Fractional programming for communication systems—part
  {II}: Uplink scheduling via matching,'' \emph{IEEE Transactions on Signal
  Processing}, vol.~66, no.~10, pp. 2631--2644, 2018.

\bibitem{al}
P.-A. Absil, R.~Mahony, and R.~Sepulchre, \emph{Optimization algorithms on
  matrix manifolds}.\hskip 1em plus 0.5em minus 0.4em\relax Princeton
  University Press, 2009.

\bibitem{boumal2020introduction}
N.~Boumal, ``An introduction to optimization on smooth manifolds,''
  \emph{Available online, May}, 2020.

\bibitem{hu2020brief}
J.~Hu, X.~Liu, Z.-W. Wen, and Y.-X. Yuan, ``A brief introduction to manifold
  optimization,'' \emph{Journal of the Operations Research Society of China},
  vol.~8, no.~2, pp. 199--248, 2020.

\bibitem{cvx}
M.~Grant and S.~Boyd, ``{CVX}: Matlab software for disciplined convex
  programming, version 2.1,'' \url{http://cvxr.com/cvx}, Mar. 2014.

\end{thebibliography}

\end{document}